\documentclass[12pt]{article}
\usepackage{amssymb}
\usepackage{amsmath}
\usepackage{bm}
\usepackage[dvips]{epsfig}
\usepackage{graphicx}

\begin{document}

\title{The theory of closed 1-forms, levels of quasiperiodic functions
and transport phenomena in electron systems.}

\author{A.Ya. Maltsev$^{1}$, S.P. Novikov$^{1,2}$}

\date{
\centerline{$^{1}$ \it{L.D. Landau Institute for Theoretical Physics 
of Russian Academy of Sciences}}
\centerline{\it 142432 Chernogolovka, pr. Ak. Semenova 1A}
\centerline{$^{2}$ \it{V.A. Steklov Mathematical Institute 
of Russian Academy of Sciences}}
\centerline{\it 119991 Moscow, Gubkina str. 8}
}

\maketitle

\begin{abstract}
The paper is devoted to the applications of the theory of dynamical systems 
to the theory of transport phenomena in metals in the presence of strong 
magnetic fields. More precisely, we consider the connection between the geometry 
of the trajectories of dynamical systems, arising at the Fermi surface in the 
presence of an external magnetic field, and the behavior of the conductivity 
tensor in a metal in the limit $\, \omega_{B} \tau \rightarrow \infty $. 
The paper contains a description of the history of the question and the 
investigation of special features of such behavior in the case of the appearance 
of trajectories of the most complex type on the Fermi surface of a metal.
\end{abstract}

\section{Introduction.}
\setcounter{equation}{0}

 This paper is devoted to one of the most important applications of the theory 
of dynamical systems on manifolds, namely, the description of transport phenomena 
in electron systems in condensed matter physics. The most important in this case 
is consideration of electron transport phenomena in the theory of normal metals,
where arbitrarily complex dispersion laws that describe dynamics of electrons 
in a crystal lattice in the presence of external fields can arise. Thus, 
in particular, the application of a strong magnetic field generates an 
extremely nontrivial dynamics of electron states in a metal, determined by
features of the dispersion relation (the dependence of the energy on the 
quasimomentum) for a given metal. The features of the corresponding dynamical 
system, in their own turn, determine the transport properties of the electron 
gas in the presence of strong magnetic fields, observed experimentally.
The dynamical system, arising in the space of electron states in the presence 
of a magnetic field, can be considered here as a Hamiltonian system with 
a complex Hamiltonian $ \, \epsilon ({\bf p}) \, $ and a special Poisson 
bracket defined by applied magnetic field. Another important feature of this 
system is that its phase space represents a three-dimensional torus 
$ \, \mathbb{T}^{3} \, $ but not Euclidean space. As can be shown, the theory 
of such systems is closely related both to the theory of foliations generated 
by 1-forms on manifolds, and with the theory of quasiperiodic functions on the 
plane. Let us note also here that the relationship between the geometry of 
the trajectories of these dynamical systems and electron transport properties 
in normal metals was first discovered by the school of I.M. Lifshitz in the 1950s.

 The problem of the complete classification of the trajectories of dynamical 
systems, arising in the space of electron states in the presence of a magnetic 
field for an arbitrary dispersion law, was first set by S.P. Novikov in the 
early 1980s and then intensively studied in his topological school (S.P. Novikov,
A.V. Zorich, S.P. Tsarev, I.A. Dynnikov). The study of this problem led
to the appearance of a number of deep topological theorems that formed the 
foundation for the topological classification of different trajectories that may 
arise in systems of this type. In particular, an extremely important part of 
the classification was the description of geometric properties of stable nonclosed 
trajectories of such systems (A.V. Zorich, I.A. Dynnikov), which provides a basis 
for a description of stable nontrivial regimes of behavior of magneto-conductivity 
in normal metals. Another extremely important achievement was the discovery of
two different types of unstable open trajectories in the described systems, which 
reveal much more complicated (chaotic) behavior in comparison with the stable 
trajectories (S.P. Tsarev, I.A. Dynnikov).

 The appearance of new topological results for systems describing quasiclassical 
dynamics of electron states in crystals has played an important role in the further 
development of the theory of transport phenomena in metals in the presence of
strong magnetic fields. Thus, the investigation of galvano-magnetic phenomena 
at the presence of stable open trajectories of these systems on the Fermi surface
allowed to introduce new topological characteristics (topological quantum numbers) 
observed in the conductivity of normal metals (S.P. Novikov, A.Ya. Maltsev).
At the same time, the study of galvano-magnetic phenomena in metals that allow 
the appearance of chaotic trajectories on the Fermi surface, makes it possible 
to observe special regimes in the behavior of the magneto-conductivity in strong 
magnetic fields, which were not considered before. In addition, it should also 
be noted that the topological description of different types of trajectories 
of systems describing the dynamics of electron states in crystals provides 
opportunities for a more accurate description of the behavior of the 
magneto-conductivity from the analytical point of view (A.Ya. Maltsev).
Let us note also that the obtained results are applicable not only to 
galvano-magnetic phenomena but also to many other transport phenomena 
(electron thermal conductivity, etc.) in normal metals.

 It must be said that the physical applications of the Novikov problem 
are not limited in reality only to the theory of normal metals, but are 
related also to transport phenomena in other systems. Among such systems
one can particularly distinguish two-dimensional electron systems placed 
into artificially created quasiperiodic ``superpotentials'' and actively 
investigated in experimental physics at the present time. The Novikov problem 
can be considered here either as the problem of describing the geometry of 
trajectories of special dynamical systems (of different dimensions) or as a 
problem of describing the level lines of a quasiperiodic function on a plane 
with an arbitrary number of quasiperiods. Thus, in particular, the topological 
results obtained for systems that describe the quasiclassical dynamics of 
electron states in three-dimensional crystals also make it possible to describe
in the semiclassical approximation all the main features of transport phenomena 
in a two-dimensional electron gas in potentials with three quasiperiods 
in strong magnetic fields (orthogonal to the plane of the sample). The study
of the Novikov problem with a larger number of quasiperiods is in fact
a rather complicated problem. Nevertheless, for the present moment
there is a description of an important class of potentials with four 
quasiperiods whose non-closed level lines have a ``regular form'' 
(lie in a straight strip of a finite width and pass it through) and can 
be characterized by special topological numbers (S.P. Novikov, I.A. Dynnikov).

 In this paper we will mainly discuss applications of the dynamical systems 
theory to the theory of galvano-magnetic phenomena in metals in sufficiently 
strong magnetic fields. More precisely, we will mainly consider the behavior 
of the conductivity tensor in the presence of chaotic trajectories on the Fermi 
surface. Our results here will be based on a number of properties of such 
trajectories, established as a result of intensive study of the corresponding 
dynamical systems in the most recent time. In particular, we will show that the 
asymptotic Zorich - Kontsevich - Forni indices, defined for certain classes of 
dynamical systems on two-dimensional surfaces, have a direct relation to the 
behavior of the electric conductivity tensor in strong magnetic fields.

 In the next chapter we will give a more detailed description of the connection 
between the theory of the galvano-magnetic phenomena in metals with the theory 
of dynamical systems on manifolds. In Chapter 3 we will consider the behavior 
of the tensor of electric conductivity in the presence of complex electron 
trajectories on the Fermi surface.

\section{Theory of transport phenomena in metals and dynamical systems
on manifolds.}
\setcounter{equation}{0}

 One of the most important applications of the theory of closed 1-forms on 
two-dimensional surfaces is the investigation of the geometry of quasiclassical 
electron trajectories in metals with a complex Fermi surface in the presence of 
an external magnetic field. Such situation is due, first of all, to the 
peculiarities of the space of the electron states for a fixed conduction band, 
representing a three-dimensional torus $\, \mathbb{T}^{3} \, $ from the 
topological point of view.

 More precisely, the electron states in a single crystal are described by 
solutions of the stationary Schr\"odinger equation
\begin{equation}
\label{ShredEq}
- \, {\hbar^{2} \over 2 m} \,\, \Delta \psi  \,\,\, + \,\,\, U (x, y, z) \, \psi \,\,\, = \,\,\, 
E \, \psi  \,\,\,  , 
\end{equation}
where the potential $\,\, U ({\bf r}) \, = \, U (x, y, z) \,\, $ 
is a periodic function in $\, \mathbb{R}^{3} \, $ with three independent periods 
$\, {\bf l}_{1}$, $\, {\bf l}_{2}$, $\, {\bf l}_{3}$:
$$U ({\bf r} + {\bf l}_{1}) \,\,\, \equiv \,\,\, U ({\bf r} + {\bf l}_{2}) \,\,\, \equiv \,\,\,
U ({\bf r} + {\bf l}_{3}) \,\,\, \equiv \,\,\, U ({\bf r}) $$

 The physical states of electrons in a crystal are given by bounded solutions
of the equation (\ref{ShredEq}), which can be chosen in the form of the Bloch 
functions $\, \psi_{\bf p} ({\bf r}) $, satisfying the conditions
$$\psi_{\bf p} ({\bf r} + {\bf l}_{1}) \,\,\, \equiv \,\,\, 
e^{i ({\bf p} , {\bf l}_{1}) / \hbar} \,\,
\psi_{\bf p} ({\bf r})  \,\,\, ,  \quad
\psi_{\bf p} ({\bf r} + {\bf l}_{2}) \,\,\, \equiv \,\,\, 
e^{i ({\bf p} , {\bf l}_{2}) / \hbar} \,\,
\psi_{\bf p} ({\bf r})  \,\,\, ,   $$
$$\psi_{\bf p} ({\bf r} + {\bf l}_{3}) \,\,\, \equiv \,\,\, 
e^{i ({\bf p} , {\bf l}_{3}) / \hbar} \,\,
\psi_{\bf p} ({\bf r}) $$

 The value $\, {\bf p} \, = \, (p_{1}, p_{2}, p_{3}) \, $ is called the 
quasimomentum of an electron state and completely determines this state for 
a fixed allowed energy band.\footnote{In fact, real electron states have 
also a spin degeneracy, which doubles the number of states in a real crystal.
Since for us this degeneracy does not play a significant role, we will not 
consider it here in detail.}

 It is easy to see that the quantity $\, {\bf p}\, $ is defined modulo the 
vectors
\begin{equation}
\label{ReciprLattice}
m_{1} \, {\bf a}_{1} \,\,\, + \,\,\, m_{2} \, {\bf a}_{2} \,\,\, + \,\,\,
m_{3} \, {\bf a}_{3} \,\,\, ,  \quad  m_{1}, m_{2}, m_{3} \, \in \, \mathbb{Z} \,\,\, ,
\end{equation}
where the vectors $\, {\bf a}_{1} $,  $\, {\bf a}_{2} $, $\, {\bf a}_{3} \, $ 
are determined by the relations
$${\bf a}_{1} \,\,\, = \,\,\, 2 \pi \hbar \,\, 
{{\bf l}_{2} \times {\bf l}_{3} \over ({\bf l}_{1}, {\bf l}_{2}, {\bf l}_{3})} 
\,\,\, ,  \quad
{\bf a}_{2} \,\,\, = \,\,\, 2 \pi \hbar \,\, 
{{\bf l}_{3} \times {\bf l}_{1} \over ({\bf l}_{1}, {\bf l}_{2}, {\bf l}_{3})} 
\,\,\, ,  \quad
{\bf a}_{3} \,\,\, = \,\,\, 2 \pi \hbar \,\, 
{{\bf l}_{1} \times {\bf l}_{2} \over ({\bf l}_{1}, {\bf l}_{2}, {\bf l}_{3})} $$

 The vectors $\, {\bf a}_{1} $,  $\, {\bf a}_{2} $, $\, {\bf a}_{3} \, $ 
give a basis of the so-called reciprocal lattice of a crystal that determines 
the geometry of the space of the electron states. The vectors (\ref{ReciprLattice}) 
form a complete set of the reciprocal lattice vectors $\, L^{*} \, $ in the 
$\, {\bf p} $ - space.

 Any two values of $\, {\bf p}$, that differ by a reciprocal lattice vector,
determine the same physical electron state (and the same solution of the equation 
(\ref{ShredEq})) for a fixed energy band. Consequently, we can state that 
the space of the electron states for a fixed energy band represents a 
three-dimensional torus $\, \mathbb{T}^{3}$:
$$\mathbb{T}^{3} \,\,\, = \,\,\, \mathbb{R}^{3} / L^{*} \,\,\, , $$
obtained from the $\, {\bf p}$ - space by the factorization over the vectors
of the reciprocal lattice. 

 The energy value $\, E \, = \, \epsilon ({\bf p}) \, $ for a solution of
the equation (\ref{ShredEq}) in every band is completely defined by the value 
of $\, {\bf p} \, $ and represents a smooth periodic function of $\, {\bf p} \, $ 
with the periods $\, {\bf a}_{1} $,  $\, {\bf a}_{2} $, $\, {\bf a}_{3}$.
It is easy to see that this function can also be regarded as a smooth
function on the torus $\, \mathbb{T}^{3}$ defined above. In general case,
we have an infinite number of allowed energy bands, so we have
an infinite number of different smooth functions
$\, E \, = \, \epsilon_{s} ({\bf p}) \, $, $\,\, s \, = \, 1, 2, \dots $, 
that take bounded values
$$\epsilon^{min}_{s} \,\,\, \leq \,\,\, \epsilon_{s} ({\bf p}) \,\,\, \leq \,\,\,
\epsilon^{max}_{s} $$

 It must be said that in the case of a three-dimensional crystal the energy
intervals $\, [\epsilon^{min}_{s} , \epsilon^{max}_{s}] \, $, in general,
overlap with each other, so it may be more rigorous to talk about different 
branches $\, \epsilon_{s} ({\bf p}) \, $ of the energy spectrum in a crystal.
 
 The general picture of the electron structure in a normal metal can be
represented as follows:

 The electron gas is highly degenerate, so we can assume that all the electron 
states with energies less than $\, \epsilon_{F} \, $ (the Fermi energy) are
occupied and all the electron states with energies larger than
$\, \epsilon_{F} \, $ are empty. 

 The Fermi level $\, \epsilon_{F} \, $ belongs to the interval
$\, (\epsilon^{min}_{s} , \epsilon^{max}_{s}) \, $, defined by one
of the branches of the energy spectrum in the crystal (or several
such intervals), so that the equation
\begin{equation}
\label{FermiSurface}
\epsilon_{s} ({\bf p}) \,\,\, = \,\,\, \epsilon_{F}
\end{equation}
defines a two-dimensional periodic surface in the
$\, {\bf p}$ - space. The surface defined by equation
(\ref{FermiSurface}) is called the Fermi surface of a metal.

 It is easy to see that under the condition
$$\nabla \epsilon_{s} |_{\epsilon_{F}} \,\,\, \neq \,\,\, 0 $$
the equation (\ref{FermiSurface}) defines a smooth closed surface in 
$\, \mathbb{T}^{3} \, $ after the factorization over the reciprocal 
lattice vectors. We note here that we do not require that the surface,
given by the equation (\ref{FermiSurface}) in $\, \mathbb{T}^{3} \, $,
was connected. Thus, for many normal metals the Fermi surface consists 
of several connected components. For us, however, it will be important 
that different connected components of the Fermi surface do not intersect 
each other. Let us note that the last property, as a rule, is observed 
also in the case when the full Fermi surface is given by the union of the  
surfaces (\ref{FermiSurface}) for several branches of the energy spectrum.
In the latter case, however, there can be exceptions to this rule
due to presence of symmetries of a special type in crystal lattice.
Here we will always assume that the Fermi surface is given by a set of 
smooth disjoint components embedded in $\, \mathbb{T}^{3} \, $.

 With more precise consideration, taking into account the finite 
temperatures, the described above electron distribution with respect 
to the electron states must be replaced by a temperature-dependent 
distribution function having the form:
\begin{equation}
\label{DistrFunct}
n ({\bf p}) \,\,\, = \,\,\, 1 \left/ \left( e^{(\epsilon ({\bf p}) - \epsilon_{F}) / T}
\, + \, 1 \right) \right.
\end{equation}

 In real metals, however, as a rule, we have the relation
$\, T \, \ll \, \epsilon_{F} \, $, so that the function (\ref{DistrFunct})
undergoes substantial changes only near the surface (\ref{FermiSurface}). 

 In the presence of external (constant) electric and magnetic
fields the evolution of the electron states within one energy
band can be described by a quasiclassical system describing
the change of the quasimomentum $\, {\bf p} \, $ with time
(see e.g. \cite{Abrikosov,Kittel,Ziman})
$${\dot {\bf p}} \,\,\, = \,\,\, {e \over c} \, 
\left[ {\bf v}_{gr} \times {\bf B} \right] \,\, + \,\, e \, {\bf E} \,\,\, = \,\,\,
{e \over c} \, \left[ \nabla \epsilon ({\bf p}) \times {\bf B} \right] 
\,\, + \,\, e \, {\bf E} $$ 

 When describing galvanomagnetic phenomena in strong magnetic fields
the electric field is assumed to be small, while the value of 
$\, B \, $ must satisfy the condition of the geometric limit 
$\,\, \omega_{B} \tau \, \gg \, 1 \, $, where
$\,\omega_{B} \, $ is the cyclotron frequency and
$\, \tau \, $ is the mean free time of electrons in the metal.
The magnitude of the electric field can be considered in this case 
as a small correction giving a small perturbation of the system
\begin{equation}
\label{MFSyst}
{\dot {\bf p}} \,\,\, = \,\,\, 
{e \over c} \, \left[ \nabla \epsilon ({\bf p}) \times {\bf B} \right]
\end{equation}
for the evolution of the electron states in the presence of an external
magnetic field.

 The system (\ref{MFSyst}) is an analytically integrable system
in the $\, {\bf p}$ - space whose trajectories are given by the 
intersections of surfaces of constant energy
$\,\, \epsilon ({\bf p}) \,= \, {\rm const} \,\, $ with planes,
orthogonal to the magnetic field. It is not difficult to see, however, 
that this circumstance does not allow to obtain a direct description 
of the global geometry of trajectories of (\ref{MFSyst}), since the 
$\, {\bf p}$ - space is not compact, while in the torus 
$\, \mathbb{T}^{3} \, $ the energy $\, \epsilon ({\bf p}) \, $
gives the only single-valued integral of the system (\ref{MFSyst}). 
The geometry of the trajectories of (\ref{MFSyst}), thus, 
is determined by levels of the 1-form $\, (d {\bf p} , {\bf B}) \, $,
restricted to the compact two-dimensional surfaces
$\,\, \epsilon ({\bf p}) \,= \, {\rm const} \,\, $ in the torus
$\, \mathbb{T}^{3} \, $.

 It is also easy to see that the system (\ref{MFSyst}) preserves
also the volume element $\, d^{3} p \, $, which in fact is a consequence 
of the Hamiltonian property of this system with respect to the Poisson
bracket
$$\{ p_{1} \, , \, p_{2} \} \,\, = \,\, {e \over c} \, B^{3} \,\,\, , \quad
\{ p_{2} \, , \, p_{3} \} \,\, = \,\, {e \over c} \, B^{1} \,\,\, , \quad
\{ p_{3} \, , \, p_{1} \} \,\, = \,\, {e \over c} \, B^{2} $$ 

 As a consequence of this fact, the dynamical system (\ref{MFSyst})
does not change the equilibrium electron distribution. Nevertheless,
the behavior of trajectories of system (\ref{MFSyst}) on the Fermi 
surface has a significant effect on the behavior of conductivity in strong 
magnetic fields, determined by the linear response of the electron system 
to a small electric field.

\begin{figure}[t]
\begin{center}
\includegraphics[width=0.9\linewidth]{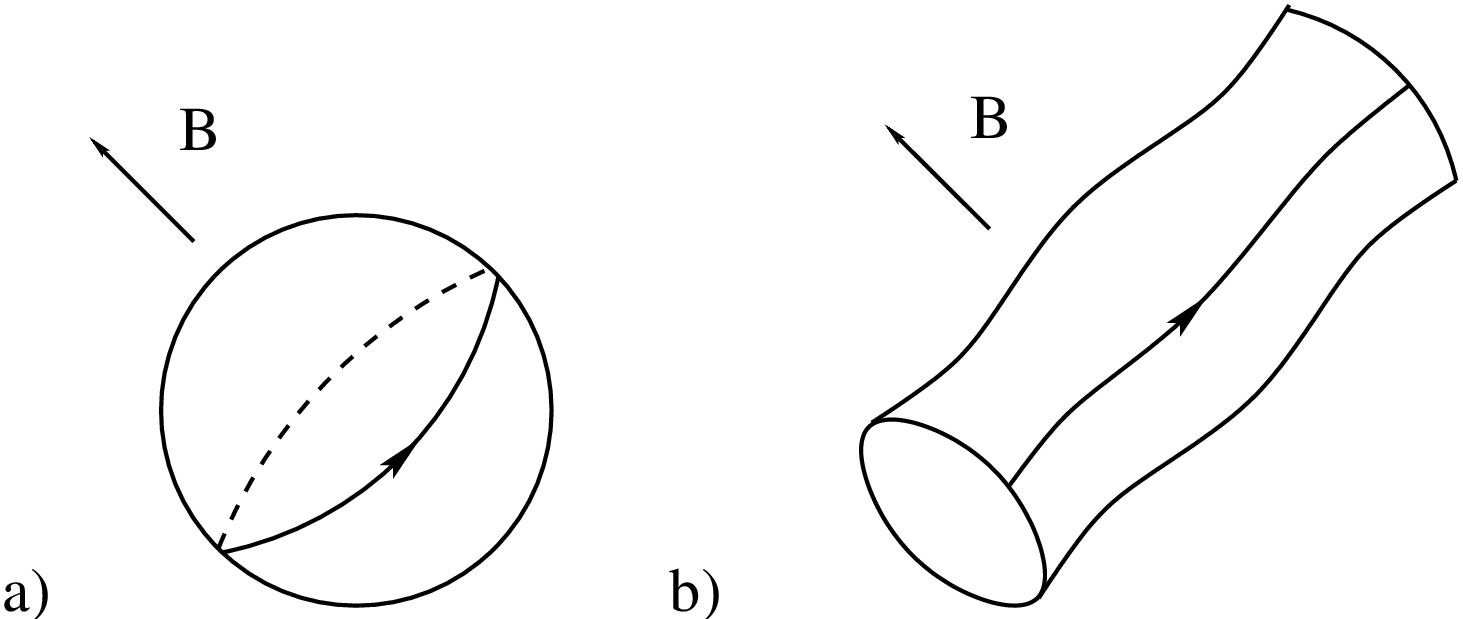}
\end{center}
\caption{The Fermi surfaces, containing only closed (a)
and open periodic trajectories (b) in the
$ \, {\bf p}$ - space.}
\label{DiffSurf}
\end{figure}

 The influence of the geometry of trajectories of system (\ref{MFSyst})
on the behavior of conductivity in strong magnetic fields was first 
discovered by the school of I.M. Lifshitz in the 1950s. Thus, in the work 
\cite{lifazkag} an essentially different behavior of conductivity
in the plane orthogonal to $\, {\bf B} \, $ in the case when
the Fermi surface contains only closed trajectories
(Fig. \ref{DiffSurf}, a) and in the case when it contains open
periodic trajectories in the $\, {\bf p}$ - space
(Fig. \ref{DiffSurf}, b) was indicated.

 Let us agree here that the coordinates in the physical space
are chosen in such a way that the axis $\, z \, $ is directed along
the magnetic field $\, {\bf B} \, $. In this case, the asymptotic
behavior of the conductivity tensor in strong magnetic fields
in the case of presence of closed trajectories only can be expressed 
by the formula
\begin{equation}
\label{Closed}
\sigma^{ik} \,\,\,\, \simeq \,\,\,\,
{n e^{2} \tau \over m^{*}} \, \left(
\begin{array}{ccc}
( \omega_{B} \tau )^{-2}  &  ( \omega_{B} \tau )^{-1}  &
( \omega_{B} \tau )^{-1}  \cr
( \omega_{B} \tau )^{-1}  &  ( \omega_{B} \tau )^{-2}  &
( \omega_{B} \tau )^{-1}  \cr
( \omega_{B} \tau )^{-1}  &  ( \omega_{B} \tau )^{-1}  &  *
\end{array}  \right)  ,   \quad \quad
\omega_{B} \tau \,\, \rightarrow \,\, \infty 
\end{equation}

 The value $\, n \,$ represents here the concentration of the 
conductivity electrons and $\, m^{*} \, $ represents the effective
mass of the electron in the crystal. The value $\, \omega_{B} \, $ 
is defined by the relation $\, \omega_{B} \, = \, e B / m^{*} c \, $.
Here and everywhere we use the symbol $\, * \, $  to denote just some 
dimensionless constant of the order of unity.

 In the case of the presence of open periodic trajectories on the 
Fermi surface it is convenient to choose the $\, x $ - axis in the 
direction of the periodic trajectories in the $\, {\bf p}$ - space.
Note, that the quasiclassical trajectories in the $\, {\bf x}$ - space 
have such form that their projections onto the plane orthogonal to 
$\, {\bf B} \, $ can be obtained from the trajectories in the 
$\, {\bf p}$ - space by rotation to $90^{\circ}$. The asymptotic 
behavior of the conductivity tensor in strong magnetic fields is 
expressed in this case by the formula
$$\sigma^{ik} \,\,\,\, \simeq \,\,\,\,
{n e^{2} \tau \over m^{*}} \, \left(
\begin{array}{ccc}
( \omega_{B} \tau )^{-2}  &  ( \omega_{B} \tau )^{-1}  &
( \omega_{B} \tau )^{-1}  \cr
( \omega_{B} \tau )^{-1}  &  *  &  *  \cr
( \omega_{B} \tau )^{-1}  &  *  &  *
\end{array}  \right)  ,   \quad \quad
\omega_{B} \tau \,\, \rightarrow \,\, \infty $$

 As can be easily seen, in the second case the conductivity is 
characterized by a strong anisotropy in the plane orthogonal to 
$\, {\bf B} \, $, which allows to determine the mean direction 
of the open trajectories experimentally.

 In the works \cite{lifpes1,lifpes2} open trajectories of
more general form on the Fermi surfaces of different shapes 
were considered. The trajectories studied in \cite{lifpes1,lifpes2} 
are not periodic, nevertheless, they also have a mean
direction in $\, {\bf p}$ - space, which also leads
to a strong anisotropy of the conductivity tensor in the plane,
orthogonal to $\, {\bf B} \, $.

 The works \cite{lifkag1,lifkag2,lifkag3}, and also the book \cite{etm},
give a review of a wide range of questions of the electron theory of 
metals, and, in particular, of questions related to the geometry of 
open trajectories of the system (\ref{MFSyst}), considered in that 
period. We would also like to refer here to the work 
\cite{KaganovPeschansky}, containing a return to this class of problems,
which also includes consideration of aspects that arose in a later 
period. 

 The general problem of describing the geometry of open  trajectories 
of the system (\ref{MFSyst}) with an arbitrary dispersion law 
$\, \epsilon ({\bf p}) \, $ was set by S.P. Novikov
(\cite{MultValAnMorseTheory}) and was actively explored in his
topological school in recent decades. The detailed study of the system 
(\ref{MFSyst}) led to the full classification of possible types 
of trajectories of this system, which allowed also to describe all 
possible types of asymptotic regimes of behavior of electric conductivity 
in strong magnetic fields.

 The most important part of the classification of open trajectories
of system (\ref{MFSyst}) is a description of stable open trajectories 
of this system that have remarkable topological properties.
These properties of open trajectories of system (\ref{MFSyst}) are 
determined by the topology of the carriers of such trajectories on the 
level surfaces $\,\, \epsilon ({\bf p}) \, = \,{\rm const} \,\, $
(in particular, on the Fermi surface), which can be defined as connected 
components of such a surface, obtained after removing of all closed 
trajectories of the system (\ref{MFSyst}). We would like to note here 
that the main results in this direction were obtained in the works 
\cite{zorich1,dynn1} which provide a basis for the description of stable 
open trajectories of the system (\ref{MFSyst}).

 In particular, in the work \cite{zorich1} the following important
statement about open trajectories of the system (\ref{MFSyst})
on generic Fermi surfaces was proved:
 
 For any rational direction of $\, {\bf B} \, $ there exists 
a neighborhood of this direction such that for any direction 
of $\, {\bf B} \, $ from this neighborhood the carriers of 
open trajectories of the system (\ref{MFSyst}) on the level 
surface $\,\, \epsilon ({\bf p}) \, = \, \epsilon_{0} \,\, $
(if they exist) represent two-dimensional tori
$\, \mathbb{T}^{2} \, $ (possibly with holes cut out),
embedded in $\, \mathbb{T}^{3} \, $.

 In the work \cite{dynn1} a similar assertion was proved for open 
trajectories of the system (\ref{MFSyst}), which are stable with
respect to variations of the energy level $\, \epsilon_{0} \, $.
Thus, according to \cite{dynn1}, if open trajectories of the system
(\ref{MFSyst}) exist at energy levels
$\,\, \epsilon ({\bf p}) \, = \, {\rm const} \,\, $ in some
neighborhood of the energy value $\, \epsilon_{0} \, $, then the 
carriers of such trajectories are also two-dimensional tori
$\, \mathbb{T}^{2} \, $ (possibly with holes cut out),
embedded in the torus $\, \mathbb{T}^{3} \, $.

\begin{figure}[t]
\begin{center}
\includegraphics[width=0.9\linewidth]{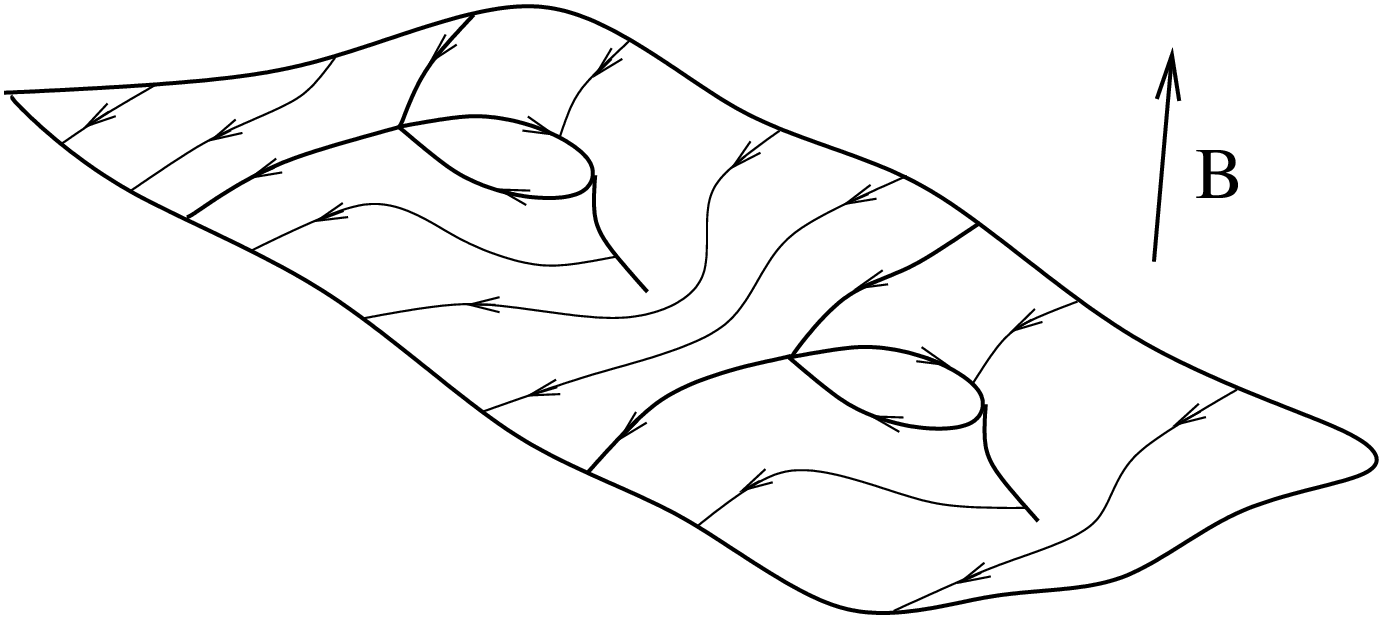}
\end{center}
\caption{A carrier of topologically regular open trajectories
in $\, {\bf p}$ - space.}
\label{Support}
\end{figure}

 The covering carriers of open trajectories in the 
$\, {\bf p}$ - space represent periodically deformed integral 
(i.e., generated by two vectors of the reciprocal lattice) planes 
with flat holes, orthogonal to $\, {\bf B} \, $ (Fig. \ref{Support}).
The corresponding open trajectories are given in this case by
intersections of such planes with planes orthogonal to the
magnetic field and represent quasiperiodic curves in the
$\, {\bf p}$ - space. In both above cases for generic directions 
of $\, {\bf B} \, $, the two-dimensional
tori $\, \mathbb{T}^{2} \, $, and also their coverings, are
locally stable with respect to small rotations of the
direction of $\, {\bf B} \, $ and with respect to small
variations of the value $\, \epsilon_{0} \, $.

\begin{figure}[t]
\begin{center}
\includegraphics[width=0.9\linewidth]{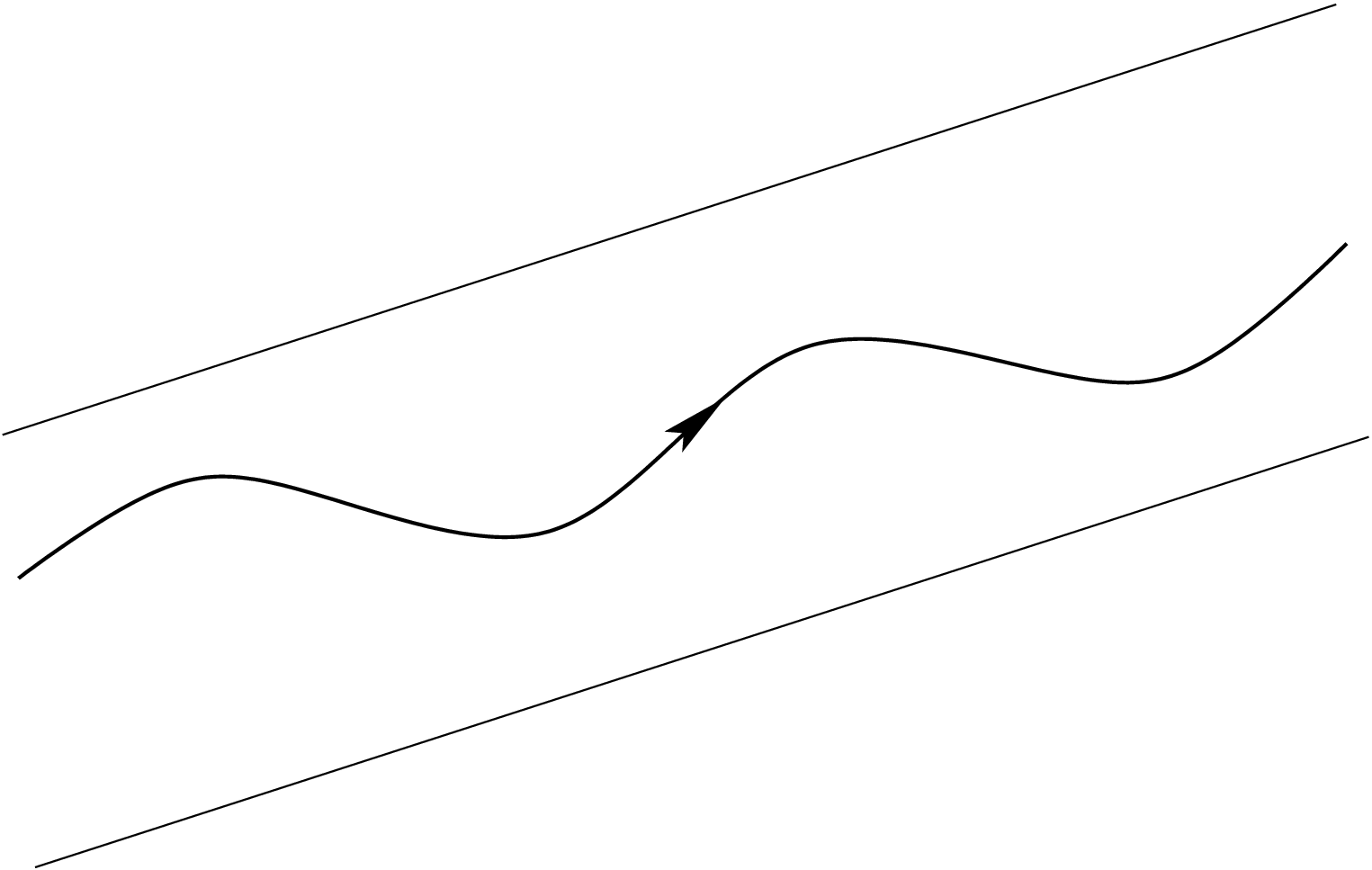}
\end{center}
\caption{A topologically regular trajectory lying
in a strip of a finite width and passing through it.}
\label{StrPol}
\end{figure}

 Based on the above description of stable open trajectories
of system (\ref{MFSyst}), we can specify the following properties 
of such trajectories which play an important role in the theory of 
galvanomagnetic phenomena in metals:

\vspace{1mm}

1) Every stable open trajectory in $\, {\bf p}$ - space lies in a straight 
strip of a finite width in the plane, orthogonal to  $\, {\bf B} \, $,
passing through it (\cite{dynn1992}) (Fig. \ref{StrPol});
 
\vspace{1mm} 
 
2) The mean direction of the open trajectories in the
$\, {\bf p}$ - space is given by the intersection of the plane,
orthogonal to $\, {\bf B} \, $, with some integral plane, which is 
the same for a fixed ``Stability Zone'' in the space of directions
of $\, {\bf B} \, $.

\vspace{1mm}

 Let us note here that the property (1) was originally
formulated by S.P. Novikov in the form of a conjecture,
which, thus, was proved for stable open trajectories in the 
works \cite{zorich1}, \cite{dynn1992} and \cite{dynn1}.
Note also that, as follows from \cite{zorich1} and \cite{dynn1}, 
it is enough to require the stability of trajectories with 
respect to either small rotations of the direction of
$\, {\bf B} \, $ or to small variations of the energy value
$\, \epsilon_{0} \, $.
 
 Properties (1) and (2) of stable (regular) open
trajectories of the system (\ref{MFSyst}) are extremely
important in the theory of galvanomagnetic phenomena in normal
metals and served as a basis for the introduction in 
\cite{PismaZhETF} of important topological characteristics 
of the electron spectrum in a metal observable in the study
of the electric conductivity. Thus, the presence of the 
property (1) for stable open trajectories leads to strong anisotropy
of conductivity in the plane orthogonal to $\, {\bf B} \, $,
observable experimentally in the limit
$\, \omega_{B} \tau \, \gg \, 1 \, $. 
The direction of the greatest suppression of conductivity
(in $\, {\bf x}$ - space) coincides with the mean direction 
of the open trajectories in $\, {\bf p}$ - space and is, therefore,
observable experimentally. The property (2), together with the
property of local stability of carriers of open
trajectories, allows us to determine the homological class
of the embedding
$\,\, \mathbb{T}^{2} \, \subset \, \mathbb{T}^{3} \, $,
given by an irreducible triple of integers
$\, (m_{1}, m_{2}, m_{3}) \, $. The numbers
$\, (m_{1}, m_{2}, m_{3}) \, $  were named in \cite{PismaZhETF}
the topological quantum numbers observable in the conductivity of 
normal metals and represent (together with the form of the 
corresponding Stability Zone in the space of directions of
$\, {\bf B} $) important characteristics of the electron spectrum 
in a metal.

 Let us note here that the topological quantum numbers characterize
geometric properties of the electric conductivity tensor and are
well-observable quantities in the experiment. At the same time,
determination of the shape of a Stability Zone may require more
precise experimental studies, in particular, the experimentally 
observable Stability Zone may differ from the exact mathematical 
Stability Zone. We must also say, that the analytic properties of 
the conductivity tensor in the experimentally observed Stability Zone 
can also be quite nontrivial (see e.g. \cite{AnProp,CyclRes}).

 It is interesting to note that the theorem of \cite{zorich1}
admits a generalization to the case of functions on the plane with 
four quasiperiods. In this formulation trajectories of the system 
(\ref{MFSyst}) in the $\, {\bf p}$ - space can be considered as 
the level lines of a function on the plane having three quasiperiods, 
i.e. the function obtained by restriction of a periodic function 
in three-dimensional space on a plane embedded in
$\, \mathbb{R}^{3} \, $. Similarly, one can consider the level lines 
of functions on the plane having four quasiperiods, i.e. functions 
obtained by restriction of a 4-periodic function on the plane embedded 
in $\,\mathbb{R}^{4} \, $. It is natural to call here the plane
integral if it is generated by two independent periods of the 
corresponding periodic function in $\,\mathbb{R}^{4} \, $.
As can be shown (see \cite{NovKvazFunc, DynNov}), for 
noncompact level lines of generic quasiperiodic functions on the 
plane with four quasiperiods the following assertion is true:

 For any integral embedding of a two-dimensional plane in
a four-dimensional space there is an open neighborhood
in the space of the directions of embedding
$\,\,\mathbb{R}^{2} \,\subset \, \mathbb{R}^{4} \, $,
such that all the noncompact level lines of the corresponding
quasiperiodic functions on the plane

1) lie in straight strips of finite widths, passing through them;

2) have a mean direction defined by the intersection
of the embedded plane with some integral three-dimensional
plane in $\,\mathbb{R}^{4} \, $, which is constant for this
neighborhood.

 Let us also note here that the description of the geometry of 
the level lines of quasiperiodic functions on the plane can play
important role in the theory of transport phenomena in two-dimensional
electron systems in the presence of an external ``super-potential'', 
studied in modern experiments (see e.g. \cite{JMathPhys}).

 For a description of the general picture that appears on the angular 
diagram, which describes the conductivity behavior as a function of the 
direction of $\, {\bf B} \, $, it is convenient to use the general 
description of behavior of open trajectories of (\ref{MFSyst}), arising 
at all energy levels $\,\, \epsilon ({\bf p}) \,=\, {\rm const} \,\, $
for a fixed energy band (\cite{dynn3}).

 According to \cite{dynn3}, open trajectories of system (\ref{MFSyst})
for a fixed direction of  $\, {\bf B} \, $ always arise either
in a connected energy interval
$\,\, \epsilon_{1} ({\bf B}) \, \leq \, \epsilon \, \leq \, \epsilon_{2} ({\bf B}) \, $,
or only at one energy level
$\,\, \epsilon \, = \, \epsilon_{1} ({\bf B}) \, = \, \epsilon_{2} ({\bf B}) \, $.
In the first case, the open trajectories of (\ref{MFSyst}) have topologically
regular form described above and the corresponding direction of 
$\, {\bf B} \, $ belongs in generic case to some Stability Zone defined
for the entire energy spectrum $\,\epsilon ({\bf p}) \, $.
The corresponding Stability Zones represent regions with piecewise smooth 
boundaries on the unit sphere $\, \mathbb{S}^{2}\, $ and form an everywhere 
dense set in the space of directions of $\, {\bf B} \, $.
As shown in \cite{dynn3}, the unit sphere $\, \mathbb{S}^{2}\, $ 
can consist either of one Stability Zone for a certain type of the spectrum 
$\, \epsilon ({\bf p}) \, $, or contain infinitely many such Zones.
Within each Stability Zone the mean direction of regular open trajectories 
of the system (\ref{MFSyst}) in the $\, {\bf p}$ - space is given by the 
intersection of the plane orthogonal to $\, {\bf B} \, $ with some integral
plane, which is the same for a given Zone. We also note here that the practical 
calculation of the boundaries of the Stability Zones and the corresponding 
topological quantum numbers represents a non-trivial computational problem 
based on serious topological methods (see, e.g. \cite{DeLeoPhysB}).

\vspace{1mm}

 In the case of the existence of open trajectories of system (\ref{MFSyst})
just at one energy level such trajectories can be characterized by both 
topologically regular and more complex chaotic behavior.

\vspace{1mm}

 Coming back to the theory of galvanomagnetic phenomena in metals, 
we must now fix an energy level in the conduction band, given by the 
Fermi energy, and consider the behavior of the trajectories of system
(\ref{MFSyst}) only at this level. It is not difficult to see here that
in this case open trajectories are present at the Fermi level either if
the Fermi energy belongs to the interval
$\, [ \epsilon_{1} ({\bf p}) , \epsilon_{2} ({\bf p}) ] \, $ or
coincides with a single energy level containing open
trajectories for a given direction of $\, {\bf B} \, $.
As a consequence, the Stability Zones for a fixed Fermi energy do not 
form anymore an everywhere dense set on the unit sphere $\,\mathbb{S}^{2}\, $
and in general we have also regions, corresponding to the presence 
of only closed trajectories on the Fermi surface, at the angular diagram.
In addition to the Stability Zones and the regions corresponding to the 
presence of only closed trajectories, the angular diagram can contain 
also special directions of $\, {\bf B} \, $, corresponding to chaotic 
behavior of open trajectories of system (\ref{MFSyst}) on the Fermi surface.

 The above picture gives a general description of the typical angular
diagram for the magneto-conductivity of a metal with a complex Fermi 
surface. It must be said that in the general case the angular diagram can
have some additional features for special directions of the magnetic field.
The most detailed consideration of various opportunities for open trajectories 
that arise for a general dispersion law, can be found in \cite{dynn3}.
A detailed consideration of transport phenomena connected with different 
behavior of the open trajectories of system (\ref{MFSyst})
was presented in the papers \cite{UFN,BullBrazMathSoc,JournStatPhys}.
Let us also note here that the angular diagram associated with the full
dispersion law is possibly also available for the experimental research in 
semiconductor structures in very strong magnetic fields (see \cite{JETP1}).

\section{The chaotic trajectories and the
conductivity behavior in strong magnetic fields.}
\setcounter{equation}{0}

 We want to stop now on a more detailed description of 
chaotic behavior of the trajectories of the system (\ref{MFSyst}), 
which is also possible in the case of complex Fermi surfaces.
As we have already said above, such trajectories can exist only 
on one energy level for a given direction of $\, {\bf B} \, $.
We note at once that chaotic trajectories can naturally be divided 
into two different types (Tsarev type and Dynnikov type) corresponding 
to different ``degrees of irrationality'' of the direction of the magnetic 
field. Thus, the trajectories of the Tsarev type arise in the case of
``partially irrational'' directions of $\, {\bf B} \, $, such that
the plane, orthogonal to $\, {\bf B} \, $,  contains a reciprocal
lattice vector. The Tsarev trajectory (\cite{Tsarev}) was the first 
example of a trajectory characterized by explicit chaotic behavior on
the Fermi surface in $\,\mathbb{T}^{3} \, $. In the planes,
orthogonal to $\, {\bf B} \, $ in the covering $\, {\bf p}$ - space,
trajectories of the Tsarev type can not be enclosed in a straight line
of a finite width and, from this point of view, also are not 
topologically regular. Nevertheless, the trajectories of the Tsarev type
possess asymptotic direction in planes orthogonal to $\, {\bf B} \, $,
which also leads to strong anisotropy of the conductivity tensor
in strong magnetic fields. It can be shown (see \cite{dynn2}) that
the last property always takes place for chaotic trajectories
in the case of ``partially irrational'' directions of $\, {\bf B} \, $.
 
\begin{figure}[t]
\begin{center}
\includegraphics[width=0.9\linewidth]{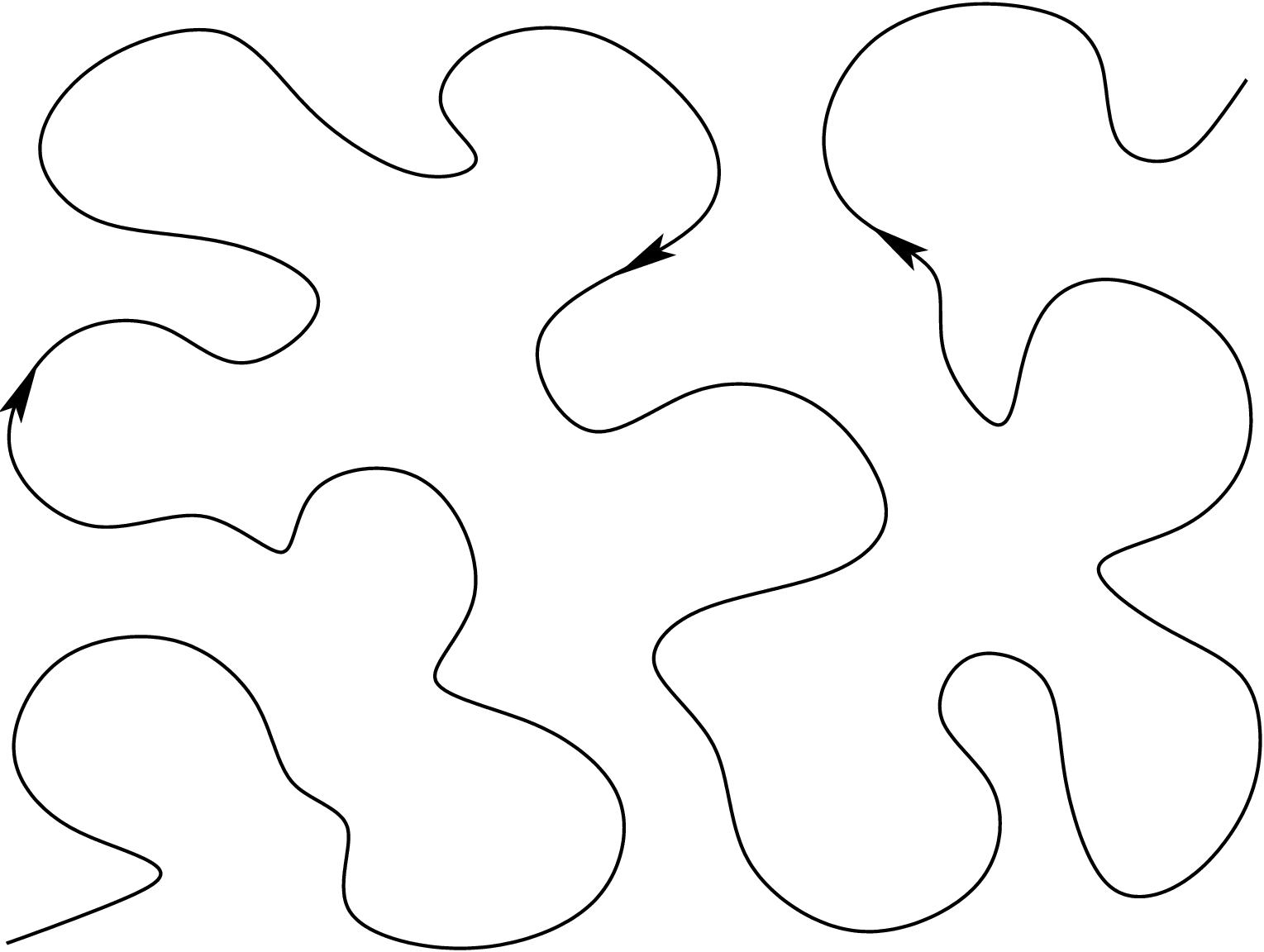}
\end{center}
\caption{Schematic view of a Dynnikov chaotic trajectory
in the plane orthogonal to $\, {\bf B} \, $.}
\label{Complex}
\end{figure}

 The chaotic trajectories of Dynnikov type arise only for directions of
$\, {\bf B} \, $ having maximal irrationality (the plane orthogonal 
to $\, {\bf B} \, $ does not contain reciprocal lattice vectors).
The trajectories of this type are characterized by an explicit chaotic 
behavior both on the Fermi surface in $\,\mathbb{T}^{3} \, $ and in 
the covering $\, {\bf p}$ - space (Fig. \ref{Complex}). As a rule, 
each carrier of such trajectories represents a surface of genus 3 
embedded in $\,\mathbb{T}^{3} \, $ with the covering of the 
``maximal rank'' in the $\, {\bf p}$ - space. More precisely, 
the carriers of such trajectories (obtained after removal
of closed trajectories from the Fermi surface) form closed
surfaces of genus 3 (or more) after gluing the holes, arising
after the removal of the cylinders of closed trajectories, by flat 
discs, orthogonal to $\, {\bf B} \, $. Let us also note that compact
piecewise smooth surfaces obtained in this way can be regularized by 
an arbitrarily small perturbation to smooth surfaces,
embedded in $\,\mathbb{T}^{3} \, $. It is easy to see also that
the corresponding regularization of the system (\ref{MFSyst}) does not 
change geometric properties of open trajectories in $\, {\bf p}$ - space
on a large scale (\cite{dynn3}).

 The set of directions of the magnetic field corresponding to the appearance
of chaotic trajectories of the Dynnikov type at any of the energy levels
represents a rather complex set (of the Cantor type) on the angular
diagram $\, \mathbb{S}^{2} \, $. According to the conjecture of S.P. Novikov,
the Hausdorff dimension of its subset on  $\, \mathbb{S}^{2} \, $,
corresponding to a fixed Fermi surface of general position, is strictly 
less than 1. Let us also note here that different properties of the 
trajectories of this type are actively investigated in modern works
(see e.g. \cite{AvilaHubSkrip1, AvilaHubSkrip2, DeLeo1, DeLeo2, DeLeo3,
DeLeoDynnikov1, DeLeoDynnikov2, DeLeo2017, DynnSkrip1, DynnSkrip2, Skripchenko1, 
Skripchenko2, zorich2}). The chaotic trajectories arising for directions of 
$\, {\bf B} \, $ of maximal irrationality are the most complex
and lead to the most nontrivial regimes in conductivity behavior 
in strong magnetic fields. 

 The behavior of the electric conductivity in strong magnetic fields 
in the presence of chaotic trajectories of Dynnikov type was considered in 
\cite{ZhETF2} and is characterized by significant differences from the 
conductivity behavior in the presence of only closed or topologically regular 
trajectories of system (\ref{MFSyst}). One of the main differences in the
conductivity behavior in this case is the suppression of the conductivity 
along the direction of the magnetic field for $\,\,\omega_{B} \tau \,\gg \,1 \,$.
Another important feature of the conductivity behavior in the presence of
trajectories of this type is the appearance of fractional powers of the 
parameter $\, \omega_{B} \tau \, $ in the asymptotics of the components 
of the conductivity tensor in the limit 
$\,\, \omega_{B} \tau \,\rightarrow \,\infty \, $.

 It should be noted that the introduction of fractional powers of the 
parameter $\, \omega_{B} \tau \, $ in \cite{ZhETF2} in the description of
the conductivity tensor was based on a special property of chaotic open 
trajectories (self-similarity) constructed in \cite{dynn2}. This property 
is connected with the behavior of these trajectories on a large scale and 
is represented by the existence of two directions in the plane orthogonal 
to $\, {\bf B} \, $ and the corresponding scale coefficients 
$\, \lambda_{1} \, $, $\, \lambda_{2} \, $, such that after stretching 
of the plane along these directions with coefficients $\, \lambda_{1} \, $,
$\, \lambda_{2} \, $ the new trajectories obtained can be reduced to
the initial form by a finite deformation in the plane. The presence of such 
a remarkable property for the trajectories constructed in \cite{dynn2}, 
leads, in particular, to the appearance of two fractional parameters 
observable in the asymptotics of the components of the conductivity tensor
in the plane orthogonal to $\, {\bf B} \, $. It must be said, however,
that such a remarkable property is not observed in the general case
for chaotic trajectories arising for directions of $\, {\bf B} \, $ of
maximal irrationality.

 Here we want to show, however, that the appearance of fractional powers
in the asymptotics of the components of the electric conductivity tensor 
is related in fact to a more general phenomenon in the theory of dynamical
systems and should, apparently, have a general character in the presence 
of open trajectories of this type on the Fermi surface.

 More precisely, we will show here that the fractional powers describing the 
asymptotic behavior of the components of the symmetric part of the conductivity 
tensor in the plane, orthogonal to $\, {\bf B} \, $, in the limit of strong 
magnetic fields have a direct relation to the Zorich - Kontsevich - Forni 
indices, defined for dynamical systems on compact surfaces 
(see \cite{zorich2,Zorich1996,ZorichAMS1997,Zorich1997,zorich3,ZorichLesHouches}).
Our consideration here will be based on the description of these characteristics,
presented in the paper \cite{zorich3}, and we will assume for simplicity
that the carriers of open trajectories on the Fermi surface have genus 3.
Let us note here, that the results of \cite{zorich3} are formulated as 
taking place for ``almost all'' foliations generated by closed 1-forms on 
compact surfaces. In particular, these results are valid for the level lines
of 1 - forms obtained from constant 1 - forms for almost any embedding 
of a two-dimensional surface $\, M^{2}_{g} \, \rightarrow \, \mathbb{T}^{N} \, $
into a torus $\, \mathbb{T}^{N} \, $ of sufficiently large dimension.
For the embeddings $\, M^{2}_{g} \, \rightarrow \, \mathbb{T}^{3} \, $,
in general, the conditions that lead to the existence of the
Zorich - Kontsevich - Forni indices for chaotic trajectories on the surface
$\, M^{2}_{g} \, $, require additional justification. As an example of such 
a justification, we can point the paper \cite{AvilaHubSkrip1}, where the 
construction and investigation of chaotic trajectories on the Fermi surface 
of a sufficiently general form was carried out. We shall consider here the 
properties of the electric conductivity tensor in the limit 
$\, \omega_{B} \tau \, \rightarrow \, \infty \, $ in the presence of
chaotic trajectories on the Fermi surface, corresponding to certain 
Zorich - Kontsevich - Forni indices. We shall use here the simplest 
kinetic model of the conductivity in crystals and represent the key points
of the corresponding derivation on the ``physical'' level of rigorousness.
Nevertheless, the general form of the relations represented here 
remains true also after moving to more accurate physical models that 
take into account all the features of the electron kinetics in real
conductors. For greater convenience, we can assume that we always deal
with a regularized foliation on a smooth surface
$\,  M^{2}_{3} \, $ of genus 3 whose open fibers have the same geometric
properties as the open trajectories of the system (\ref{MFSyst}).

 Following \cite{zorich3}, we will now assume that the foliation generated 
by the 1 - form $\, (d {\bf p} , {\bf B}) \, $ on the regularized carrier
of open trajectories, has the following properties:

 Consider a layer (level line) in general position and fix on it a
starting point $\, P_{0} \, $. Fix on the same layer a point 
$\, P_{1} \, $, in which this layer comes close to $\, P_{0} \, $ after 
passing a sufficiently large path along the carrier of open trajectories.
Connect the points $\, P_{0} \, $ and $\, P_{1} \, $ by a short segment and 
define, thus, a closed cycle on the carrier of open trajectories.
Let us denote the homology class of the resulting cycle by
$\, c_{P_{0}} (l) \, $, where $\, l \, $ is the length of the corresponding 
segment of the layer in some metric.

 There is a flag of subspaces
$$V_{1} \,\, \subset \,\, V_{2} \,\, \subseteq \,\, V_{3} \,\, \subseteq \,\, 
V \,\, \subset \,\, H_{1} ( M^{2}_{3}; \mathbb{R} ) \,\,\, , $$
such that:

 1) For any such layer $\, \gamma \, $ and any point
$\,\, P_{0} \, \in \, \gamma \,\, $  
$$\lim_{l \rightarrow \infty} \,\, {c_{P_{0}} (l) \over l} \,\,\, = \,\,\, c \,\,\, , $$
where the non-zero asymptotic cycle
$\,\, c \, \in \, H_{1} ( M^{2}_{3}; \mathbb{R} ) \,\, $  is proportional 
to the Poincare cycle and generates the subspace $\, V_{1}\, $.

 2) For any linear form
$\,\,\phi \, \in \, Ann (V_{j}) \, \subset \, H_{1} ( M^{2}_{3}; \mathbb{R} ) \, $,
$\,\, \phi \notin \, Ann (V_{j+1}) \,\, $
\begin{equation}
\label{LimSupRel}
\limsup_{l \rightarrow \infty} \,\,
{\log |\langle \phi ,  c_{P_{0}} (l) \rangle | \over \log l} \,\,\, = \,\,\, \nu_{j+1}
\quad , \quad \quad j \, = \, 1, 2 
\end{equation}

 3) For any
$\,\,\phi \, \in \, Ann (V) \, \subset \, H_{1} ( M^{2}_{3}; \mathbb{R} ) \, $,
$\,\, || \phi || \, = \, 1 \,\, $  
$$|\langle \phi ,  c_{P_{0}} (l) \rangle | \,\, \leq \,\, const  \,\,\, , $$
where the constant is determined only by foliation.

 4) The subspace $\,\, V \, \subset \, H_{1} ( M^{2}_{3}; \mathbb{R} ) \,\, $
is Lagrangian in homology, where the symplectic structure
is determined by the intersection form.

 5) Convergence to all the above limits is uniform in
$\, \gamma \, $ and $\,\, P_{0} \, \in \, \gamma \, $, i.e. 
depends only on $\, l \, $. 

\vspace{1mm}

 Let us note that in generic situation we have in this case
$\,\, 1 \, > \, \nu_{2} \, > \, \nu_{3} \, > \, 0 \, $,
$\,\, dim \, V_{2} \, = \, 2 \, $, $\,\, dim \, V_{3} \, = \, dim \, V \, = \, 3 \, $. 

\vspace{1mm}

 Note also that the description presented in \cite{zorich3} is based on
a sequence of representations of the corresponding minimal component
(the carrier of chaotic trajectories) as a union of a fixed number of classes
(layers) of the cycles described above, whose lengths tend to infinity. Cycles
of one class are almost identical for each of these representations and cover
a finite area of the corresponding minimal component.

\vspace{1mm}

 For the description of the properties of the trajectories of system 
(\ref{MFSyst}) in $\, {\bf p}$ - space which we need let us consider 
the map in homology, induced by the embedding
$\,\, M^{2}_{3} \, \subset \, \mathbb{T}^{3} \, $.
It is not difficult to see that the image of each space $\, V_{j} \, $ 
belongs to the two-dimensional space defined by the plane orthogonal to
$\, {\bf B} \, $. Besides that, in examples of chaotic trajectories of 
Dynnikov type there is no linear growth of the deviation from the point 
$\, P_{0} \, $ with increasing of the length of the trajectory in the plane 
orthogonal to $\, {\bf B} \, $, which means that the image of the asymptotic 
cycle $\, c \, $ is equal to zero. The image of the space $\, V_{2} \, $ is
one-dimensional and determines a selected direction in the plane,
orthogonal to  $\, {\bf B} \, $, along which the mean deviation of a
trajectory grows faster with its length $\,( \sim l^{\nu_{2}}) \, $,
than in the orthogonal direction. In the generic case we suppose also 
that the image of the space $\, V_{3} \, $ is two-dimensional and is given 
by the plane orthogonal to $\, {\bf B} \, $. It is not hard to see that 
the 1 - forms $\, d p_{x} \, $ and $ \, d p_{y} \, $ give in this case
a necessary basis of 1 - forms, to which the statement (2) formulated above
can be applied. Therefore, we can choose a coordinate system $\, (x, y, z) \, $ 
in such a way that for some frame sequences of the values $ \, l \, $ 
we will have the relations
\begin{equation}
\label{Estim1}
| \Delta p_{x} (l) | \,\,\, \simeq \,\,\, p_{F} \, 
\left( {l \over p_{F}} \right)^{\nu_{2}} \quad , \quad \quad
| \Delta p_{y} (l) | \,\,\, \simeq \,\,\, p_{F} \,
\left( {l \over p_{F}} \right)^{\nu_{3}} 
\end{equation}
for the corresponding deviations of the trajectory along the coordinates 
$\, p_{x} \, $ and $\, p_{y} \, $ when passing a part of the above approximate
cycles on the Fermi surface. The value $\, p_{F} \, $ represents here the 
Fermi momentum, approximately equal to the size of the Brillouin zone in 
the $\, {\bf p}$ -space.

\vspace{1mm}

 Let us explain here what we really mean by the relations given above.
The existence of the relations (\ref{LimSupRel}) means the existence 
for any $\, \delta > 0 \, $ of certain frame sequences 
$\, \{ l^{\prime}_{k} \} \, $, $\, \{ l^{\prime\prime}_{k} \} \, $,
such that we have the relations
\begin{equation}
\label{AccurateSootn1}
A^{\prime}_{k} \,\, p_{F} \, \left( 
{l^{\prime}_{k} \over p_{F}} \right)^{\nu_{2} - \delta} \,\,\, < \,\,\,
| \Delta p_{x} (l) | \,\,\, \leq \,\,\, B^{\prime}_{k} \,\, p_{F} \,
\left( {l^{\prime}_{k} \over p_{F}} \right)^{\nu_{2}} \,\,\, ,
\end{equation}
\begin{equation}
\label{AccurateSootn2}
A^{\prime\prime}_{k} \,\, p_{F} \, \left( 
{l^{\prime\prime}_{k} \over p_{F}} \right)^{\nu_{3} - \delta} \,\,\, < \,\,\,
| \Delta p_{x} (l) | \,\,\, \leq \,\,\, B^{\prime\prime}_{k} \,\, p_{F} \,
\left( {l^{\prime\prime}_{k} \over p_{F}} \right)^{\nu_{3}} \,\,\, ,
\end{equation}
where the sequences $\, \{ A^{\prime}_{k} \} $, 
$\, \{ A^{\prime\prime}_{k} \} $, $\, \{ B^{\prime}_{k} \} $,
$\, \{ B^{\prime\prime}_{k} \} \, $ are bounded by any (arbitrarily small) 
power of the values $\, l_{k} \, $.

 In the experimental measurement of the quantities $\, \nu_{j} \, $, 
however, we can consider the value $\, \delta \, $ lying within the experimental 
error and not distinguish the powers $\, \nu_{j} \, $ and 
$\, \nu_{j} - \delta \, $. Thus, under the relations (\ref{Estim1}) we mean 
actually the relations (\ref{AccurateSootn1}) - (\ref{AccurateSootn2}), which we 
will write here in this abbreviated form.

\vspace{1mm}

 At the same time, for all the cycles described above, having lengths of the 
order of $ \, l \, $, we can write the estimations
$$| \Delta p_{x} (l) | \,\,\, \leq \,\,\, p_{F} \, 
\left( {l \over p_{F}} \right)^{\nu_{2}} \quad , \quad \quad
| \Delta p_{y} (l) | \,\,\, \leq \,\,\, p_{F} \,
\left( {l \over p_{F}} \right)^{\nu_{3}} $$
 
 We note here that the cycles, corresponding to the estimate (\ref{Estim1}), 
sweep out a finite area on the full Fermi surface for each reference value 
$ \, l \, $, which can represent arbitrary fraction of the full area of
the Fermi surface given by any number from 0 to 1 for each of these values.
It can be noted here that the latter circumstance is also unimportant in the 
experimental determination of the quantities $\, \nu_{j} \, $ due to the 
so-called ``logarithmic insensitivity'' of these quantities to the size of 
the area covered by the corresponding sections of the trajectories.
 
 Directly from the system (\ref{MFSyst}), we can then write (for the same 
reference values of $\, l$) the estimations
\begin{equation}
\label{IntV}
\left| \oint \, v^{x}_{gr} \left( t \right) \, d t \right| 
\,\,\, \simeq \,\,\, {c p_{F} \over e B} \, \left( {l \over p_{F}}
\right)^{\nu_{3}} \quad \quad , \quad 
\left| \oint \, v^{y}_{gr} \left( t \right) \, d t \right| 
\,\,\, \simeq \,\,\, {c p_{F} \over e B} \, \left( {l \over p_{F}}
\right)^{\nu_{2}} 
\end{equation}
for the same part of the described cycles and
$$\left| \oint \, v^{x}_{gr} \left( t \right) \, d t \right| 
\,\,\, \leq \,\,\, {c p_{F} \over e B} \, \left( {l \over p_{F}}
\right)^{\nu_{3}} \quad \quad , \quad 
\left| \oint \, v^{y}_{gr} \left( t \right) \, d t \right| 
\,\,\, \leq \,\,\, {c p_{F} \over e B} \, \left( {l \over p_{F}}
\right)^{\nu_{2}} $$
for the rest of the cycles.

\vspace{1mm}

 Coming back to the theory of galvanomagnetic phenomena in metals, we must consider
the kinetic equation for the distribution function $\, f ({\bf p}, t) \, $
\begin{equation}
\label{KinEq}
f_{t} \,\,\, + \,\,\, {e \over c} \, \sum_{k=1}^{3} \,
\left[ \nabla \epsilon ({\bf p}) \times {\bf B} \right]^{k} \,\,
{\partial f \over \partial p^{k}} \,\,\, + \,\,\, e \, \sum_{k=1}^{3} \,
E^{k} \, {\partial f \over \partial p^{k}} 
\,\,\,\, =  \,\,\,\, I [f] ({\bf p}, \, t)
\end{equation}
in the presence of external electric and magnetic fields. We will be interested
here in stationary solutions of the equation (\ref{KinEq}), so that we can put
in fact $\, f({\bf p}, \, t) \, = \, f({\bf p}) $. The functional $\, I [f] \, $ 
is the collision integral responsible for the relaxation of perturbations of the 
distribution function to its equilibrium values $\, f_{0} ({\bf p}) \, $.
In particular, its properties also play a decisive role for the magnitude of 
the constant response of an electron system to a constant external influence.
In the linear approximation in $\, {\bf E} \, $ we can write the linearized equation 
for the linear response $\, f_{1} ({\bf p}) \, $: 
\begin{equation}
\label{LinKinEq}
{e \over c} \, \sum_{k=1}^{3} \,
\left[ \nabla \epsilon ({\bf p}) \times {\bf B} \right]^{k} \,\,
{\partial f_{(1)} \over \partial p^{k}} \,\,\, + \,\,\, 
e \, \sum_{k=1}^{3} \, E^{k} \, 
{\partial f_{0} \over \partial p^{k}} 
\,\,\,\, =  \,\,\,\,
\left[ {\hat L}_{[f_{0}]} \, \cdot \, f_{(1)} \right] ({\bf p}) \quad ,
\end{equation}
where $\, {\hat L}_{[f_{0}]} \, $ represents the linearization of the functional 
$\, I [f] ({\bf p}) \, $ on the corresponding function
$\, f_{0} $. In the so-called $\, \tau $ - approximation, the right-hand side of 
the system (\ref{LinKinEq}) can be replaced by the expression
$\, - f_{1} ({\bf p}) / \tau \, $, where $\, \tau \, $ plays the role of 
a characteristic relaxation time (electron mean free time). Let us note here that 
a more complicated form of the linearized collision integral does not change 
in fact the results represented below.

 It is easy to see that in the $\, \tau $ - approximation the equations on 
$\, f_{1} ({\bf p}) \, $ represent ordinary differential equations written along
trajectories of the system (\ref{MFSyst}). It can also be seen that the physical 
solutions of the corresponding equations are concentrated near the Fermi surface 
due to the corresponding behavior of the quantities
$${\partial f_{0} \over \partial p^{k}} \,\,\, = \,\,\,
{\partial f_{0} \over \partial \epsilon} \,\, v^{k}_{gr} ({\bf p}) $$ 

 As a consequence of this fact, the expression for the conductivity 
(taking into account the spin degeneracy) is actually determined by the behavior 
of the solutions of (\ref{LinKinEq}) on the Fermi surface and can be represented 
in the following general form
\begin{equation}
\label{GenSigmakl}
\sigma^{ik} (B) \,\,\,\,\, = \,\,\,\,\, 
{2 e c  \over  B}  \, \iint_{S_{F}}
\, {d p_{z} \, d s \over (2 \pi \hbar)^{3}} \,\,\,
v^{i}_{gr} (p_{z}, s) \,\, \int_{-\infty}^{s} \,
v^{k}_{gr} (p_{z}, s^{\prime}) \,\,\,
e^{{c (s^{\prime} - s) \over e B \tau}} \,\, d s^{\prime} 
\end{equation}
where $\, s \, = \, t e B / c \, $ is the Hamiltonian parameter along the 
trajectories of system (\ref{MFSyst}).

 The contribution of the open trajectories to the conductivity tensor 
$\, \Delta \sigma^{ik} \, $ is given here by the restriction of the integral 
in (\ref{GenSigmakl}) to the set of carriers of open trajectories 
$\, {\hat S}_{F} \, $ instead of the whole Fermi surface, so that we can write:
\begin{equation}
\label{DeltaSigmaB}
\Delta \, \sigma^{ik} (B) \,\,\,\,\, = \,\,\,\,\, 
{2 e c  \over  B}  \, \iint_{\hat{S}_{F}}
\, {d p_{z} \, d s \over (2 \pi \hbar)^{3}} \,\,\,
v^{i}_{gr} (p_{z}, s) \,\, \int_{-\infty}^{s} \,
v^{k}_{gr} (p_{z}, s^{\prime}) \,\,\,
e^{{c (s^{\prime} - s) \over e B \tau}} \,\, d s^{\prime} 
\end{equation}

 As we have said, we shall assume here that the set of carriers of open 
trajectories on the Fermi surface is represented by a single connected component 
of genus 3.

 For the contribution of the open trajectories on the Fermi surface to the 
symmetric part of the conductivity tensor it is not difficult to obtain after 
some calculations the following expression
\begin{equation}
\label{SymmetricPart}
\Delta \, s^{ik}(B) \,\,\,\, = \,\,\,\,
2 \, e^{2} \, \tau \,\, \iint_{{\hat S}_{F}} \,
\langle v^{i}_{gr} \rangle_{B} \,\,
\langle v^{k}_{gr} \rangle_{B}
\,\,\, {d p_{z} \, d s \over (2 \pi \hbar)^{3}}  \,\,\,\, , 
\end{equation}
where
\begin{equation}
\label{VAveragedB}
\langle v^{i,k}_{gr} \rangle_{B} \, (p_{z}, s)
\,\,\,\, \equiv \,\,\,\,
{c \over e B \tau} \, \int_{-\infty}^{s} \,
v^{i,k}_{gr} \,(p_{z}, s^{\prime}) \,\,\,
e^{{c (s^{\prime} - s) \over e B \tau}} \,\, d s^{\prime} 
\end{equation}

 To estimate the values (\ref{VAveragedB}), we can now use the relations 
(\ref{IntV}) if we use the approximation
\begin{equation}
\label{VAvEst}
\langle v^{i,k}_{gr} \rangle_{B} \, (p_{z}, s)
\,\,\,\, \simeq \,\,\,\, {c \over e B \tau} \,
\int_{s - e B \tau / c}^{s} \,\,\,
v^{i,k}_{gr} \,(p_{z}, s^{\prime}) \,\, d s^{\prime}  
\end{equation}
on the corresponding trajectories of the system (\ref{MFSyst}).

 It should be noted that between the Hamiltonian parameter $\, s \, $ 
and the trajectory length $\, l \, $ there may actually be differences 
due to the presence of singular points of the system (\ref{MFSyst}) on 
the Fermi surface. In our case, however, when calculating the conductivity 
containing contributions from all trajectories, this difference 
will not play a significant role in the main order. In this order, 
we can thus use the estimate $\, s \, \simeq \, m^{*} l / p_{F} \, $ as 
the common for the trajectories of the system (\ref{MFSyst}) and write
the approximations
$$\langle v^{i,k}_{gr} \rangle_{B} \, (p_{z}, s)
\,\,\,\,\, \simeq \,\,\,\,\, {1 \over \tau} \,
\int_{l \, = \, s p_{F} / m^{*} \, - \, \omega_{B} \tau p_{F}}^{l = s p_{F} / m^{*}} 
\,\,\, v^{i,k}_{gr} \, \left(p_{z}, {m^{*} l \over p_{F}} \right) \,\, d t(l) $$ 

 Analyzing the behavior of the trajectories in the plane orthogonal to 
$\, {\bf B} \, $, one can in fact see that for the corresponding reference values 
$\, \omega_{B} \tau \, = \, l / p_{F} \, $ we can now use 
(using ``logarithmic insensitivity'') the estimations
$$\langle v^{x}_{gr} \rangle_{B} \, (p_{z}, s)
\,\,\,\,\, \simeq \,\,\,\,\, {p_{F} \over m^{*}} \,\,
\left( \omega_{B} \tau \right)^{\nu_{3}-1} \quad , $$
$$\langle v^{y}_{gr} \rangle_{B} \, (p_{z}, s)
\,\,\,\,\, \simeq \,\,\,\,\, {p_{F} \over m^{*}} \,\,
\left( \omega_{B} \tau \right)^{\nu_{2}-1} $$
on some finite area of the Fermi surface and
$$\langle v^{x}_{gr} \rangle_{B} \, (p_{z}, s)
\,\,\,\,\, \leq \,\,\,\,\, {p_{F} \over m^{*}} \,\,
\left( \omega_{B} \tau \right)^{\nu_{3}-1} \quad , $$
$$\langle v^{y}_{gr} \rangle_{B} \, (p_{z}, s)
\,\,\,\,\, \leq \,\,\,\,\, {p_{F} \over m^{*}} \,\,
\left( \omega_{B} \tau \right)^{\nu_{2}-1} $$
on the remaining area.

 The above estimations permit us to write the relations
$$\Delta \, s^{xx} (B) \,\,\, = \,\,\, \Delta \sigma^{xx} (B) \,\,\, \simeq \,\,\,
{n e^{2} \tau \over m^{*}} \,
\left( \omega_{B} \tau \right)^{2\nu_{3}-2} \quad , $$
\begin{equation}
\label{Sigmadiagest}
\Delta \, s^{yy} (B) \,\,\, = \,\,\, \Delta \sigma^{yy} (B) \,\,\, \simeq \,\,\,
{n e^{2} \tau \over m^{*}} \, 
\left( \omega_{B} \tau \right)^{2\nu_{2}-2}
\end{equation}
for the same frame sequences.

\begin{figure}[t]
\begin{center}
\includegraphics[width=0.9\linewidth]{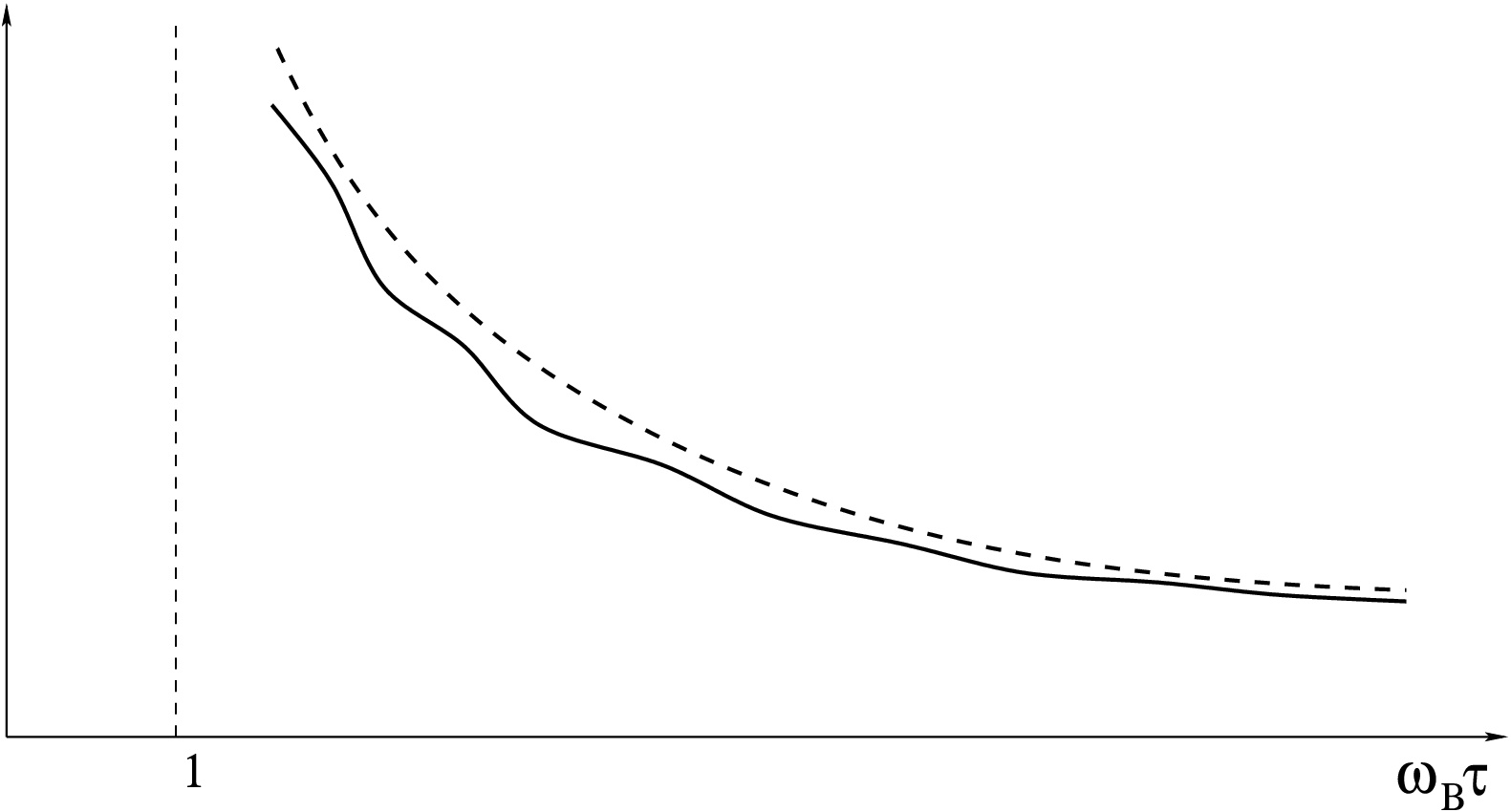}
\end{center}
\caption{The common trend and the actual behavior of the quantities
$\, \sigma^{xx} (B) \, $ and $\, \sigma^{yy} (B) \, $ in the limit
$\, \omega_{B} \tau \, \rightarrow \, \infty \, $.}
\label{Steps}
\end{figure}

 We note here that the estimations (\ref{Sigmadiagest}) do not represent the principal 
term in any asymptotic expansion of the quantities $\, \sigma^{xx} (B) \, $ and 
$\, \sigma^{yy} (B) \, $, but only characterize some ``general trend'' in the behavior 
of these quantities, which in the general case is accompanied by some fluctuations 
in their values as functions of $\, \omega_{B} \tau \, $ (Fig. \ref{Steps}).

 Strictly speaking, we also have here just the relations
$$\limsup_{\omega_{B} \tau \rightarrow \infty} \,\, 
{\log \Delta \sigma^{xx} (B) \over \log \omega_{B} \tau} 
\,\,\, = \,\,\, 2\nu_{3} \, - \, 2 \,\,\, , \quad \quad
\limsup_{\omega_{B} \tau \rightarrow \infty} \,\, 
{\log \Delta \sigma^{yy} (B) \over \log \omega_{B} \tau} 
\,\,\, = \,\,\, 2\nu_{2} \, - \, 2 $$

 Nevertheless, the regimes (\ref{Sigmadiagest}) are directly related to 
the asymptotic behavior of $\, \sigma^{xx} (B) \, $ and  $\, \sigma^{yy} (B) \, $, 
since the fluctuations in their magnitudes are actually limited by a number of 
additional conditions. In particular, $\, \sigma^{xx} (B) \, $ and 
$\, \sigma^{yy} (B) \, $ are monotone (decreasing) functions of $\, B \, $, 
which imposes significant restrictions on the deviation of their values from 
the general trend. We also note here that the replacement of the estimate 
(\ref{VAvEst}) with a more accurate definition (\ref{VAveragedB}) 
(or using the exact collision integral), in general, smoothes such oscillations 
while maintaining the presence of a common trend (\ref{Sigmadiagest}).

 Similarly, we can write the relations
$$\limsup_{\omega_{B} \tau \rightarrow \infty} \,\, 
{\log | \Delta s^{xy} (B) | \over \log \omega_{B} \tau} \,\,\, \leq \,\,\, 
\nu_{2} \, + \, \nu_{3} \, - \, 2 \,\,\, , $$
which allow us to write
$$| \Delta s^{xy} (B) | \,\,\, \leq \,\,\, 
\left( \omega_{B} \tau \right)^{\nu_{2} + \nu_{3} - 2} $$
as an estimate for the common trend for these quantities.

\vspace{1mm}

 The described above contribution of chaotic open trajectories to conductivity  
should be in general added with the contribution of short closed trajectories, 
which usually also present on the Fermi surface. From the formula (\ref{Closed}) 
it is easy to see, however, that they give a vanishingly small contribution 
in the plane orthogonal to $\, {\bf B} $, as compared to the one described above, 
in the limit $\, \omega_{B} \tau \, \rightarrow \, \infty $.

\vspace{1mm}

 We can see that in the presence of the Dynnikov chaotic trajectories, the 
conductivity in the plane orthogonal to $\, {\bf B} \, $ also reveals anisotropy, 
although it is not expressed so much as in the case of stable open trajectories.
We also note that the picture described above does not change significantly if 
we assume that the minimal components (carriers) containing chaotic trajectories 
have genus greater than 3 and the corresponding dynamics are also described by 
some Zorich - Kontsevich - Forni indices. 

 Indeed, in this case, we must similarly assume existence of a flag of subspaces
$$V_{1} \,\, \subset \,\, V_{2} \,\, \subset \,\, \dots \,\, \subset \,\, 
V_{g} \,\, \subset \,\, V \,\, \subset \,\,
H_{1} ( M^{2}_{g}; \mathbb{R} ) \,\,\, , $$ 
possessing the properties (1)-(5) presented above, and we can also assume
in the generic case 
$\,\, 1 \, > \, \nu_{2} \, > \, \nu_{3} \, > \, \dots \, \nu_{g} \, > \, 0 \, $,
$\,\, dim \, V_{2} \, = \, 2 \, $, $\,\, dim \, V_{3} \, = \, 3$,
$\,\, \dots \, $, $\,\, dim \, V_{g} \,= \, g \, $. Generically,
we can now assume that the image of $\, V_{2} \, $ under the embedding
$\, M^{2}_{g} \, \rightarrow \, \mathbb{T}^{3} \, $ is one-dimensional
and the images of $\, V_{3} \, $, $ \, \dots $, $\, V_{g} \, $ coincide with 
the plane orthogonal to $\, {\bf B} \, $. Repeating exactly the arguments given 
above, it is easy to see that the asymptotic behavior of the components of 
the symmetric part of the conductivity tensor in the plane orthogonal to 
$\, {\bf B} \, $ is now determined by the largest indices $\, \nu_{2} \, $ 
and $\, \nu_{3} \, $ and the image of the space $\, V_{2} \, $ under the 
embedding $\, M^{2}_{g} \, \rightarrow \, \mathbb{T}^{3} \, $.

\vspace{1mm}

 Let us make here one more remark about chaotic trajectories of Dynnikov type 
and the properties of the electric conductivity tensor in strong magnetic fields.
As it can in fact be shown, the chaotic trajectories arising for directions of 
$\, {\bf B} \, $ of the maximal irrationality can in some sense be divided into 
two different classes (see \cite{dynn4,Skripchenko2,DynnSkrip1,DynnSkrip2}). 
Namely, let us consider a fixed plane in $\, {\bf p}$ -space containing chaotic 
trajectories and ask the question: how many connected components (i.e. how many 
different chaotic trajectories) is contained in a given plane?
It turns out that only two situations are possible:

\vspace{1mm}

1) Almost all planes, orthogonal to $\, {\bf B} \, $, contain only one chaotic 
trajectory;

\vspace{1mm}

2) Almost all planes, orthogonal to $\, {\bf B} \, $, contain infinitely many 
chaotic trajectories.

\vspace{1mm}

 It is not difficult to see that the belonging of a dynamic system to one of 
the above classes should definitely affect the geometric properties of the 
corresponding chaotic trajectories on large scales. Thus, in the first case 
the trajectory densely sweeps out the whole plane and it is possible to estimate 
the area, covered by a segment of the trajectory of length $\, l \, $, 
by the value of $\, l \, $. In the second case, each of the chaotic trajectories 
sweeps the plane much less densely, leaving large areas for other trajectories 
in the same plane. In the first case, therefore, we can expect the presence of 
the special relation $\, \nu_{2} \, + \, \nu_{3} \, = \, 1 \, $
for the corresponding Zorich - Kontsevich - Forni indices, while in the second 
case one should expect inequality $\, \nu_{2} \, + \, \nu_{3} \, > \, 1 \, $.
Coming back to the asymptotics of the components of the conductivity tensor, 
we can see that the second case corresponds to a slower decreasing of the 
conductivity in different directions in the plane, orthogonal to $\, {\bf B}$, 
as the magnetic field increases.

\vspace{1mm}

 In conclusion, let us make one more remark about the behavior of the tensor
of electric conductivity in strong magnetic fields in the presence of chaotic
trajectories on the Fermi surface. In a sense, the reasoning we have given is 
of a theoretical nature and the question of experimental investigation of the 
corresponding regimes must be studied separately. In particular, we can ask
about applicability of the physical model of the electron transport used 
above and possible corrections to the regimes, described by us, in real 
crystals. Here we would like to indicate the main of the quantum corrections 
to the quasiclassical consideration presented above, which substantially 
affects the conductivity behavior in strong magnetic fields. Namely, we would 
like to mention the phenomenon of (intraband) magnetic breakdown, which should 
be observed on chaotic trajectories in a sufficiently strong magnetic field.

\begin{figure}[t]
\begin{center}
\includegraphics[width=0.9\linewidth]{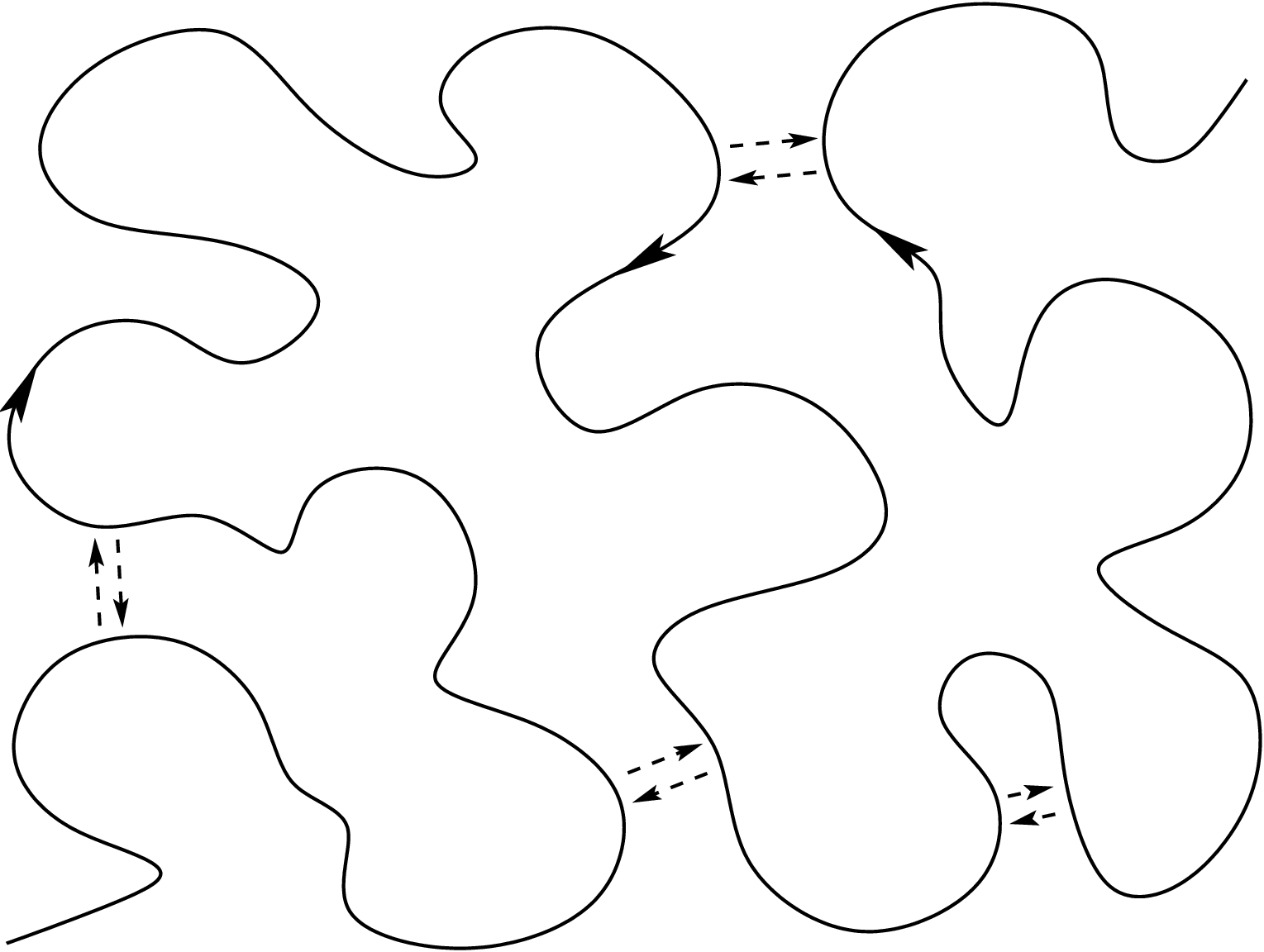}
\end{center}
\caption{The phenomenon of magnetic breakdown on a chaotic trajectory 
in the $\, {\bf p}$ - space.}
\label{Proboi}
\end{figure}

 The magnetic breakdown is a quantum phenomenon that allows an electron to 
``jump'' from one section of a quasiclassical trajectory to another preserving 
the value $\, p_{z} \, $ if the corresponding sections of the trajectory approach 
sufficiently close to each other in the $\, {\bf p}$ - space (Fig. \ref{Proboi}).
The probability of a quantum jump is higher, the closer to each other the 
corresponding sections of the trajectory and the larger the value of $\, B \, $.
In the limit of small distances between the sections or very large values of 
$\, B \, $ the corresponding probability tends to $\, 1/2 $. Arguing very 
approximately, for a given value of $\, B \, $ we can introduce the 
characteristic distance $\, \delta p (B) \, $, such that the magnetic 
breakdown can be considered improbable at a distance between the sections
$\, \Delta p \, > \, \delta p (B) \, $ and quite probable for 
$\, \Delta p \, < \, \delta p (B) \, $. We consider here very schematically 
the effect of intraband breakdown on the behavior of the components of
$\, \sigma^{ik} (B) \, $ in our case.

 It is easy to see that, because the carriers of chaotic trajectories 
necessarily contain saddle singular points, on each of the Dynnikov 
chaotic trajectories we have sections that are arbitrarily close to other 
parts of the trajectory (or sections of other trajectories) separated by 
segments of trajectories of different lengths. For a given value of the 
magnetic field, we can introduce the characteristic length $\, l^{(1)}_{B} \, $
of the trajectory in the $\, {\bf p}$ - space that separates two sections that 
approach at the distance $\, \leq \delta p (B) \, $ to other sections of the
trajectory ($l^{(1)}_{B} \, $ decreases with the growth of $\, B $).
At the same time, we can introduce the characteristic length $\, l^{(2)}_{B} \, $
passed by the electron along a trajectory in the $\, {\bf p}$ - space between two 
acts of scattering by impurities ($l^{(2)}_{B} \, \sim \, B \tau $).
It is easy to see that the intraband magnetic breakdown has small effect on 
the quasiclassical regimes of conductivity that we describe, if we have the 
condition $\, l^{(1)}_{B} \, \gg \, l^{(2)}_{B} \, $. At the same time, under 
the condition $\, l^{(1)}_{B} \, \leq \, l^{(2)}_{B} \, $ the magnetic breakdown 
should have a significant effect on the conductivity behavior. We note that the 
quantities $\, l^{(1)}_{B} \, $ and $\, l^{(2)}_{B} \, $ depend on features 
of the electron spectrum and should be evaluated separately in each specific case.
It is not difficult to see here that for a rough estimate of the effect of the
magnetic breakdown on the regimes described by us it is possible to use the 
change of the time $\, \tau \, $ to the effective electron mean free time 
$\, \tau_{eff}(B) \, $, using the formula 
$$\tau^{-1}_{eff} \,\,\, = \,\,\, \tau^{-1} \, + \, \tau^{-1}_{(1)} 
\,\,\, , $$
where $\, \tau_{(1)}(B) \, \sim \, l^{(1)}_{B} / B \, $ 
plays the role of the characteristic time of the electron motion 
over the length $\,\, \simeq l^{(1)}_{B} \, $ along the trajectory.
We note here that there is also a sufficiently large number of more 
complicated quantum corrections to the conductivity, but they play 
smaller role in our situation.

\end{document}